\pdfoutput=1
\documentclass[sigplan,screen,nonacm]{acmart}

\usepackage{booktabs}
\usepackage{multirow}
\usepackage{enumitem}
\usepackage{graphicx}
\usepackage{xcolor}
\usepackage{listings}
\usepackage{pgfplots}
\pgfplotsset{compat=1.18}

\newcommand{\sys}{MasuGate}
\newcommand{\systx}{\textsc{MasuGate-Tx}}
\newcommand{\sysres}{\textsc{MasuGate-Res}}
\newcommand{\syshold}{\textsc{MasuGate-Hold}}
\newcommand{\psslong}{policy-state serializability}
\newcommand{\pss}{PSS}
\newcommand{\statelesspolicy}{stateless policy}
\newcommand{\statefulpolicy}{stateful policy}

\newcommand{\Req}{\mathsf{req}}
\newcommand{\Prin}{\mathsf{principal}}
\newcommand{\Act}{\mathsf{action}}
\newcommand{\Args}{\mathsf{args}}
\newcommand{\State}{\mathsf{S}}
\newcommand{\View}{\mathsf{Q}}
\newcommand{\Policy}{\mathsf{P}}
\newcommand{\Effect}{\mathsf{E}}
\newcommand{\Decision}{\mathsf{d}}
\newcommand{\History}{\mathsf{H}}
\newcommand{\Serial}{\mathsf{serial}}
\newcommand{\Allow}{\mathsf{allow}}
\newcommand{\Deny}{\mathsf{deny}}
\newcommand{\Rpolicy}{R^{P}}
\newcommand{\Reffect}{R^{E}}
\newcommand{\Weffect}{W^{E}}
\newcommand{\Scopes}{\mathsf{scopes}}

\lstdefinelanguage{MasuGatePolicy}{
  morekeywords={policy,on,allow,deny,escalate,when,otherwise,and,or,not,in},
  morekeywords=[2]{args,context,principal,accounts,ledger,approvals,risk,budget},
  morecomment=[l]{\#},
  sensitive=true,
}
\lstdefinestyle{masugatepolicy}{
  language=MasuGatePolicy,
  basicstyle=\ttfamily\footnotesize,
  keywordstyle=\bfseries\color{blue!60!black},
  keywordstyle=[2]\color{teal!50!black},
  commentstyle=\itshape\color{black!55},
  stringstyle=\color{orange!70!black},
  columns=fullflexible,
  keepspaces=true,
  showstringspaces=false,
  breaklines=true,
  frame=single,
  framerule=0.2pt,
  rulecolor=\color{black!25},
  backgroundcolor=\color{black!2},
  xleftmargin=0.25em,
  xrightmargin=0.25em,
}

\lstdefinestyle{masugatepython}{
  language=Python,
  basicstyle=\ttfamily\footnotesize,
  keywordstyle=\bfseries\color{blue!60!black},
  commentstyle=\itshape\color{black!55},
  stringstyle=\color{orange!70!black},
  columns=fullflexible,
  keepspaces=true,
  showstringspaces=false,
  breaklines=true,
  frame=single,
  framerule=0.2pt,
  rulecolor=\color{black!25},
  backgroundcolor=\color{black!2},
  xleftmargin=0.25em,
  xrightmargin=0.25em,
}

\newtheorem{definition}{Definition}

\newtheorem{theorem}{Theorem}

\setcopyright{none}
\renewcommand\footnotetextcopyrightpermission[1]{}

\title{Stateful Governance for Concurrent Agentic Systems}

\author{Yuxiang Peng}
\affiliation{%
  \institution{Purdue University}
  \city{West Lafayette}
  \state{Indiana}
  \country{USA}
}

\author{Xiaodi Wu}
\affiliation{%
  \institution{University of Maryland, College Park}
  \city{College Park}
  \state{Maryland}
  \country{USA}
}

\begin{abstract}
AI agents are moving from advisory interfaces into systems that execute consequential operations: issuing refunds, reserving scarce inventory, provisioning cloud resources, and initiating financial transfers.
These workflows require governance over effects, not only over model outputs.
Existing safeguards often decide whether an action is allowed from the information available when the action is requested.
For stateful policies, that request-time view may be incomplete: budgets, inventory, approval status, and risk signals can change before the effect occurs, making an earlier authorization or approval stale.

This paper studies stateful governance for concurrent agentic systems.
We identify stale authorization as the core failure mode and define \emph{policy-state serializability}, a correctness condition requiring committed effects to be explainable as authorized against the policy state immediately before they occur.
We present \sys{}, a runtime architecture that keeps policies as reviewable programs while coordinating the state and effects needed to preserve their decisions.
In experiments with a PostgreSQL-backed prototype of \sys{}, the system prevents stale authorizations missed by baselines that pass policy state as ordinary request context, preserves delayed approvals while unrelated work proceeds, keeps policy evolution mostly in policy text rather than trusted provider code, and avoids policy violations in a scripted, LLM-free procurement workflow where agent-governance baselines produce stale authorizations over shared budgets and inventory.
More broadly, \sys{} suggests a path for integrating stateful governance boundaries into agent frameworks and provider-backed domains where agents act on shared resources.

\end{abstract}

\begin{document}

\maketitle

\section{Introduction}
\label{sec:introduction}

AI agents are moving from advisory chat interfaces into systems that execute operations~\cite{openaiOperator2025,microsoftAutonomousAgents2024,awsBedrockAgents2026}.
Customer-support agents issue refunds, travel agents reserve seats and rooms, infrastructure agents provision cloud resources, and finance agents initiate transfers.
These workflows invoke tools that mutate databases, send messages, and trigger durable effects that are difficult to undo.
As agent frameworks make such operations easier to compose, the safety problem shifts from filtering text to governing effects.

AI governance techniques span many layers.
Deployments use prompt instructions, safety evaluations, monitoring, audit logs, sandboxing, and human review to shape or inspect agent behavior~\cite{nistAIRMF2023,openaiSafety2026,openaiOperator2025}.
These mechanisms are valuable, but much of their assurance is empirical or procedural: evaluations may reveal failures, monitors may flag them, and reviewers may catch them, but the mechanism does not itself define the state and effect invariant that must hold when an operation commits.
Prompt engineering is the simplest example.
A prompt can ask an agent to respect a budget or wait for approval, but it does not create an enforcement boundary around the effect or the state that justifies it.

\begin{figure}[t]
  \centering
  \includegraphics[width=\linewidth]{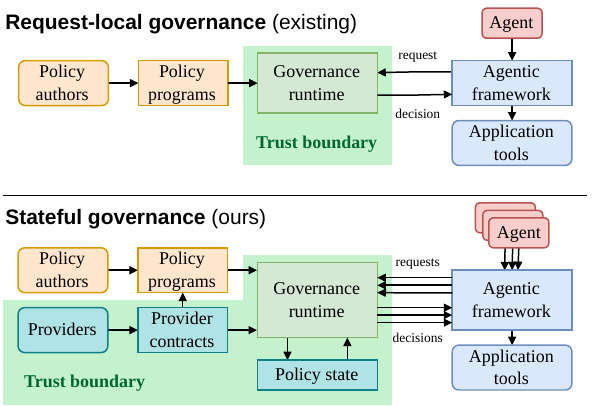}
  \caption{Request-local and stateful governance. In the request-local configuration studied here, a policy engine returns a decision before the provider applies the later effect. \sys{} adds provider contracts and certified policy state inside the trusted boundary so the runtime can protect the decision and governed effect together.}
  \label{fig:governance-schema}
  \vspace{-0.2cm}
\end{figure}

Policy-based governance is the stronger abstraction.
A policy engine evaluates explicit rules and returns a structured decision, giving a request-level decision according to the policy and the information available at that boundary.
Systems such as Cedar~\cite{cedar2024}, Microsoft's Agent Governance Toolkit (AGT)~\cite{agt2026}, and Omnigent~\cite{omnigent2026} illustrate this direction: they move governance out of prompts and ordinary application code into reviewable policy artifacts.
As Figure~\ref{fig:governance-schema} shows, a common policy-engine integration evaluates supplied request context at the request boundary and returns a decision to the agentic framework.
We use this request-local pattern as a comparison point; Section~\ref{sec:evaluation} describes the exact version-pinned baseline configurations.
This separation makes policies easier to inspect, deploy, and revise, and it is the abstraction we want to preserve.

The limitation is that this request-local boundary works best when policies check the current request: the principal, action, resource, and arguments.
This works well for access-control-like checks, but it leaves many agentic safeguards outside the policy abstraction.
Refund governance, for example, must account for a customer's recent refund history and whether an order has already been reimbursed; travel and scheduling governance must account for current inventory, existing holds, and trip budget before a seat, room, or appointment slot is reserved.
Other domains impose similar stateful conditions, including live cloud quotas, security-access risk signals, and rolling financial limits.
These policies are not simply richer checks over a single request: they require mutable governance facts maintained by the systems whose effects the agent invokes.
Statefulness is therefore necessary for expressive agent governance, not merely an implementation detail.

A stateful governance system must first satisfy two core requirements:
\begin{enumerate}[leftmargin=1.55em,itemsep=0.15ex,topsep=0.3ex,parsep=0pt,label=\arabic*.]
  \item \emph{Correctness.} It should prevent policy-violating commits: a governed action must not commit unless its policy decision remains justified under the policy semantics.
  \item \emph{Efficiency.} It should allow governed actions to proceed concurrently whenever possible.
\end{enumerate}
It should also preserve two practical design goals:
\begin{enumerate}[leftmargin=1.55em,itemsep=0.15ex,topsep=0.3ex,parsep=0pt,label=\alph*.]
  \item \emph{Human review.} It should keep an approval meaningful while a person is deciding.
  \item \emph{Modularity.} It should let policy authors change governance rules without rewriting each tool implementation.
\end{enumerate}

The difficulty is that these goals interact.
A policy decision may be correct when it is made, but unsafe by the time the governed effect occurs.
This failure appears in minimal form with two agents transferring from different accounts in the same team.
Both transfers may be individually allowed when the team has spent 9 budget credits under a 10-credit daily limit; if both 1-credit effects commit, the team has spent 11 credits.
The account writes are disjoint, but the operations conflict through policy state, the mutable state that policy evaluation depends on.
We call this failure \emph{stale authorization}: the system acts on an allow decision after the state that justified it has changed.

Human approval stretches the same problem over time: an approval may be correct when requested but invalid by resolution time after other operations consume the relevant budget, inventory, quota, or risk allowance.
The result is a \emph{stale approval}, where the human decision no longer authorizes the effect unless the system preserves or revalidates the state that justified it.

Existing approaches satisfy these goals only partially.
Request-local policy engines preserve modularity but leave the caller to protect the state behind an allow decision; global locks preserve correctness but sacrifice concurrent progress and delayed review; and hand-written transactions enforce one fixed policy but bury it in trusted tool code.
\sys{} combines these elements in a reusable stateful governance boundary between policy-engine decisions and governed effects.

We present \sys{}, a stateful governed-action runtime architecture built around an explicit provider contract.\footnote{We use \emph{contract} in the systems sense: a provider-supplied specification that tells the runtime which certified policy-state views and governed effects an operation may use, and which logical scopes must be protected.}
As shown in Figure~\ref{fig:governance-schema}, \sys{} shifts from a standalone request-local decision to a stateful boundary: policy authors write bounded stateful policies over certified policy-state views, providers declare the governed effects and mechanisms that maintain the relevant state, and the coordinator connects the two before an effect commits.
Thus, transactions and reservations remain mechanisms underneath the abstraction, while policies remain first-class governance artifacts.

With this boundary in place, we can state the desired correctness property.
We define \emph{\psslong{}} (\pss{}), a correctness property for concurrent governed operations.
Informally, every execution should have the same policy meaning as some serial execution in which each allowed effect is authorized against the policy state immediately before it is applied.
This property captures the intended decision, effect, and audit semantics for stateful governance; \sys{} is designed to enforce it under the contract assumptions stated later.

We evaluate a PostgreSQL-backed research prototype of \sys{} against the desiderata above.
The experiments test whether \sys{} prevents stale authorization when baselines lack state and effect coupling (correctness), and measure the cost of preserving that guarantee while allowing concurrent progress (efficiency).
They also study delayed approval and approval invalidation (human review), evaluate whether policy evolution remains mostly outside trusted provider code (modularity), and run a scripted, LLM-free agentic procurement benchmark against AGT and Omnigent.
Together, these controlled workloads exercise the requirements that motivate the system.

This paper makes the following contributions:

\begin{itemize}[leftmargin=1.35em,itemsep=0.15ex,topsep=0.3ex,parsep=0pt]
  \item It formalizes stale authorization caused by concurrent mutation of policy state between decision and effect, with stale approval as its long-running human-review form, and distinguishes stateless policies from policies over mutable policy state.
  \item It defines \pss{} and the contract soundness obligation needed to enforce it.
  \item It presents \sys{}, a runtime architecture that turns policy-view and effect contracts into scoped enforcement while keeping policy programs separate from trusted provider code.
  \item It implements a Python prototype of \sys{} with PostgreSQL and SQLite backends plus a benchmark harness.
  \item It evaluates \sys{} against correctness, efficiency, human review, and modularity desiderata, including a multi-policy agentic procurement workflow benchmark.
\end{itemize}

\paragraph{Positioning.}
\sys{} does not introduce a new concurrency-control primitive.
Its contribution is to make existing mechanisms usable at the governance boundary: policy authors write reviewable stateful policies, providers declare the policy-state dependencies those policies and effects touch, and the runtime protects the decision-effect interval before a governed effect commits.
The same boundary suggests a deployment path for agentic frameworks such as Microsoft Agent Framework (MAF)~\cite{microsoftAgentFramework2026} and LangGraph~\cite{langGraph2026}, where governed operations in refunds, reservations, cloud provisioning, access control, and financial transfers can pass through \sys{} without replacing the surrounding orchestration.
Section~\ref{sec:related} discusses related work in more detail.

\paragraph{Organization.}
The rest of the paper develops the problem (Section~\ref{sec:problem}), model (Section~\ref{sec:model}), design (Section~\ref{sec:design}), evaluation (Section~\ref{sec:evaluation}), limitations (Section~\ref{sec:discussion}), related work (Section~\ref{sec:related}), and conclusion (Section~\ref{sec:conclusion}).

\section{Stateful Governance}
\label{sec:problem}

This section motivates stateful governance from the policy-as-program abstraction used by modern authorization systems.
It distinguishes stateless policies from policies over mutable policy state, then uses the running budget example, request-context enforcement, and two correctness-preserving baselines to isolate stale authorization as the central consistency problem.

\subsection{Stateless and Stateful Policy as Programs}
\label{subsec:stateless-stateful}

Modern authorization systems often represent policy as a separately authored program evaluated over an authorization request~\cite{openPolicyAgent2026}.
Systems such as Cedar~\cite{cedar2024} let policy authors specify predicates over principals, actions, resources, and attributes, while application code implements the effects performed after an allow decision.
This separation between policies and effects supports independent policy review and evolution, and it is the abstraction boundary that \sys{} preserves.

In many deployments, these policies are \statelesspolicy{}.
A stateless policy depends only on information supplied with the request, such as the principal, action, resource, and attributes.
The first policy in Figure~\ref{fig:stateless-stateful-policies} has this form: it denies a transfer whose requested amount exceeds a 10-credit per-transfer cap.
Because the rule reads only \texttt{args.amount}, the runtime can evaluate it without consulting policy state such as prior transfers or shared budget usage.

\begin{figure}[t]
\centering
\begin{minipage}{0.96\linewidth}
\small
\textbf{Stateless policy}
\begin{lstlisting}[style=masugatepolicy]
deny transfer when
  args.amount > 10;
\end{lstlisting}

\small
\textbf{Stateful policy}
\begin{lstlisting}[style=masugatepolicy]
deny transfer when
  budget.spent(principal.team, 24h) + args.amount > budget.limit(principal.team);
\end{lstlisting}
\end{minipage}
\caption{Both policies govern the same transfer action. The stateless policy enforces a per-transfer cap using request-local arguments, while the stateful policy reads mutable budget state through certified policy-state views.}
\label{fig:stateless-stateful-policies}
\end{figure}

Agentic workflows put pressure on this request-local model.
Agents call tools that create durable effects, and their safeguards often depend on what has already happened.
A stateless policy can cap the current transfer, but it cannot itself ask how much the team has spent so far unless the application supplies that fact as context.
When mutable state is supplied as request context, the policy engine sees a value but does not necessarily control the underlying state or the updates that may invalidate it.
The need for stateful inputs is already visible in deployed agent tooling: AGT provides a native per-agent \texttt{CostGuard}, and Omnigent ships native cost-budget policies that read cumulative session or per-user daily spend and return \textsc{ASK} or \textsc{DENY} at request or tool-call phases~\cite{agt2026,omnigent2026}.
They both provide native stateful cost-governance mechanisms; maintaining the validity of decisions over other provider state through effect commit depends on how those mechanisms are integrated with the provider's concurrency and transaction mechanisms.

We therefore use \statefulpolicy{} for a policy program that reads mutable policy state maintained by the governed system.
The second policy in Figure~\ref{fig:stateless-stateful-policies} is stateful: it calls certified policy-state views for the team's accumulated spending and limit.
Those views are read-only from the policy program's perspective, but their values are maintained as governed effects commit.
We call the trusted component that exposes such views and realizes the corresponding governed effects a \emph{policy-state provider}.
From a systems perspective, the provider defines the boundary between policy evaluation and effect execution: policies read provider-certified state views, while the provider commits the corresponding state transitions with governed effects.

The budget policy is only one instance of this pattern.
Similar state appears in refund and payment governance, where policies depend on refund history, order status, cumulative spend, vendor status, or duplicate-payment checks; in travel, scheduling, and inventory governance, where policies depend on availability, holds, scarce inventory, and trip budget; and in cloud, security, approval, and compliance workflows, where policies depend on quotas, incident state, privileged-session state, approval basis, risk signals, or aggregate usage~\cite{taubench2024,agentdojo2024,toolemu2023,openaiOperator2025,awsBedrockAgents2026}.
Across these domains, state can encode consumable capacity, history-dependent limits, pending approvals, exclusive claims, or newly added risk signals.
The cases differ in storage and enforcement mechanism, but they share the same contract requirement: the provider must expose the policy-visible fact as a certified read-only view, map it to logical scopes, and update, validate, or reserve those scopes consistently with governed effects.

This extra expressiveness creates the core systems problem in concurrent agentic systems.
Agents may issue operations at the same time.
The research challenge is to preserve the meaning of a stateful decision from policy evaluation through effect commit.
If another operation changes the relevant policy state in between, the system can produce stale authorization.

\subsection{Stale Authorization}
\label{subsec:stateless-insufficient}

\begin{figure}[t]
\centering
\includegraphics[width=\linewidth]{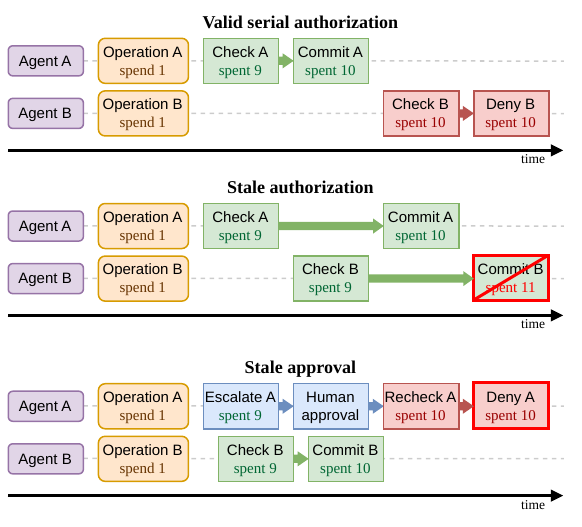}
\caption{Running budget example. Top: valid serial authorization checks B after A commits, so B is denied at 10 spent credits. Middle: stale authorization lets both checks use the same 9-credit state, so both commit and spending reaches 11. Bottom: stale approval is the delayed form: A escalates at 9 spent credits, B commits while review is pending, and A's resolution-time revalidation sees B in the current rolling window and denies A rather than committing from the old approval.}
\label{fig:stale-authorization}
\end{figure}

The transfer example below is the minimal form of the agentic workflow problem: two operations have disjoint application effects but conflict through shared policy state.
Given a stateful rule such as the budget policy above, a natural workaround is to keep the policy engine stateless and compute the stateful input outside it.
The application can attach the current aggregate to the request, and the policy can compare that supplied value with the requested action.
This preserves the surface form of a policy check, but it does not tell the runtime how the supplied value is coupled to the effect that may update it.
A minimal request-context implementation has exactly this gap:

\begin{lstlisting}[style=masugatepython]
spent = budget.spent(team, window)
decision = policy_engine.check(req, {"spent_24h": spent})
if decision == "allow":
    db.commit_transfer(sender, receiver, amount)
    budget.add_spend(team, amount)
\end{lstlisting}

The anomaly that remains is stale authorization, a check-then-act anomaly over policy state~\cite{bishop1996checking}.
Figure~\ref{fig:stale-authorization} shows three schedules for the running budget policy: valid serial authorization, stale authorization during concurrent execution, and stale approval during delayed human review.

Each check in the stale authorization schedule may be correct against the state it observes.
The violation arises because the allow decisions are not coupled to the policy-state updates performed by the effects.
In a request-context implementation, both requests may carry the same computed \texttt{spent\_24h} value, and the policy engine cannot see that the first committed transfer invalidates the second request's context.

The application-level writes are disjoint: one operation debits Alice and the other debits Bob.
The conflict is therefore invisible if the system considers only application write sets.
Both operations consume the same logical governance state, the team's rolling spending window, so correctness requires the check and effect windows to have the serial meaning shown in the top panel of Figure~\ref{fig:stale-authorization}, or an equivalent execution that validates, consumes, or reserves the relevant policy state before commit.

Human intervention stretches the same gap over a longer interval, as shown in the stale-approval panel.
A policy may escalate an operation for approval, wait while other operations commit, and then resume.
If no reservation preserves the policy-state basis of that approval, the operation must be rechecked at resolution and may be denied when the state has changed; otherwise, a once-correct approval can authorize an effect after other operations have consumed the relevant budget.

\subsection{Correctness Baselines and Abstraction Limits}
\label{subsec:incomplete-baselines}

Two baselines are correct under strong assumptions, but neither provides the abstraction boundary required for stateful governance.

\paragraph{Global serialization.}
A single global lock can serialize every governed operation from policy read through effect commit.
This is correct but overbroad: operations with disjoint policy-state dependencies wait behind one another, and a pending approval either blocks unrelated work or must release the lock and revalidate later.
Correctness should therefore be scoped to the policy state an operation depends on, rather than to the entire system.

\paragraph{Hand-written transactions.}
If policy state and effects live behind one transactional provider, the provider can enforce a fixed policy inside a serializable transaction~\cite{papadimitriou1979serializability,ports2012ssi}.
This is an important database baseline, but it moves policy logic into trusted provider code.
Policy evolution may then require changes to the transaction implementation, increasing the review burden on policy changes and raising the expertise required of policy authors.
It also weakens policy modularity: the runtime cannot directly inspect which policy-state facts a policy reads, which policy version authorized an effect, or whether a policy change can be deployed without rewriting provider code.

\smallskip
The goal is therefore a middle ground: recover the correctness of serial governed execution while allowing operations whose stateful policies and effects do not conflict to proceed concurrently.

\section{Policy-State Model and Correctness}
\label{sec:model}

This section defines the policy-state model and \emph{\psslong{}} (\pss{}), the correctness property that \sys{} targets.
The model captures the correctness obligation introduced in Section~\ref{sec:problem}: policies read mutable policy state, effects change it, and audit records must justify delayed or concurrent commits.

\subsection{Policy State}
\label{subsec:policy-state}

The policy state $\State$ is the logical state needed to decide all deployed stateful policies.
It is not necessarily the entire application state or database state.
Rather, $\State$ records exactly the governance facts that policies may consult, such as consumable capacity, histories, and approvals.

This abstraction separates policy semantics from concrete storage.
One deployment may implement $\State$ as SQL tables, another as counters, and another as service-owned logs.
The model only requires that the policy state contains the information needed for policy decisions and that governed effects update the relevant parts of that state consistently with their visible outcomes.
Time-based policies are evaluated using a provider-certified evaluation time rather than a caller-supplied clock.
The initial evaluation uses the operation's admission time, while revalidation after a delay uses a fresh provider-certified time at resolution.
A rolling-window view includes provider-committed effects that occurred strictly after the beginning of the window and no later than the evaluation time.
Consequently, an effect committed while an operation is pending is visible if it remains within the rolling window when the operation is revalidated.

\subsection{Governed Operations}
\label{subsec:governed-operations}

A request is a tuple:
\[
  \Req = (\Prin, \Act, \Args)
\]
where $\Prin$ is the authenticated principal, $\Act$ is the governed action, and $\Args$ are action arguments.
A governed operation extends a request with the policy evaluation, effect outcome, and audit outcome.

For a policy and request, the certified policy-state view vector $\vec{\View}$ is a trusted, read-only projection of policy state:
\[
  \vec{\View}(\State, \Req) \rightarrow \vec v
\]
The vector $\vec v$ contains the values returned by the registered view calls used by the policy.
The dependency information needed for enforcement is a property of the policy expression and the certified policy-state view and effect contracts; it is not part of the value vector returned to the policy program.

A policy is a pure function over the request and this value vector.
For the formal model, a terminal policy decision is either allow or deny:
\[
  \Policy(\Req, \vec v) \rightarrow
  \{\Allow, \Deny\}
\]
Policies do not execute effects or mutate policy state.
They may call only registered views with declared types and boundedness.

Implementations may also support escalation for human or external review.
Escalation is an intermediate runtime outcome that later resolves, after reservation consumption or revalidation, to the terminal allow or deny decision modeled here.

An effect is a trusted operation supplied by a policy-state provider and associated with an allowed request.
For the model, we write the effect as the transition it induces on policy state:
\[
  \Effect(\State, \Req) \rightarrow \State'
\]
The provider may update additional application state, call services, or produce external outcomes.
For policy correctness, the relevant requirement is that the policy-state transition observed by future policy decisions matches the governed effect that became visible.

\subsection{Policy-State Serializability}
\label{subsec:histories}

A history $\History$ records terminal governed operations and the externally visible ordering constraints among them.
Concretely, each operation has a begin event and a terminal event; the history can be viewed as a partial order over these events, together with the operation's request, decision, and policy-relevant effect.
For the purposes of \pss{}, an operation is terminal when it reaches allow or deny.
An allowed operation contributes its governed effect to the history, while a denied operation contributes no effect.
A pending operation created by escalation is not ordered as a completed governed operation until it resolves to allow or deny through revalidation or reservation consumption.

The real-time order induced by $\History$ orders operation $i$ before operation $j$ when $i$'s terminal event precedes $j$'s begin event in the partial order.
Thus, \pss{} is strict-serializability-like over policy state: it requires serial explainability plus real-time order, but not global serialization of the application database.
Real-time order matters for governance and audit because once a governed effect is visible, later decisions should be explainable as having seen that effect; overlapping operations are instead ordered by the serialization points chosen by the enforcement mechanism.
A serial history is a history whose terminal operations do not overlap.
The target correctness condition requires concurrent histories to have the same policy meaning as some legal serial history, following the serial-history view of concurrency correctness~\cite{papadimitriou1979serializability,herlihy1990linearizability}.
Intuitively, the serial history is an explanation of what the concurrent execution meant.
The runtime need not execute operations one at a time, but after the fact each terminal decision should fit into an order where the policy was checked against the state that made the effect valid.
The definition below makes this requirement precise.

\begin{definition}[Policy-State Serializability]
\label{def:pss}
A concurrent history $\History$ of terminal governed operations satisfies \pss{} if there exists a serial history $\Serial(\History)$ such that:
1) $\Serial(\History)$ respects the real-time order of $\History$;
2) every terminal decision in $\Serial(\History)$ matches all applicable policies evaluated immediately before the operation's serial position;
3) every allowed operation applies its governed effect at that position;
4) every denied operation produces no effect; and
5) $\Serial(\History)$ produces the same governed effects as $\History$ and leaves the same final policy state.
\end{definition}

Equivalently, for each terminal operation $i$ in the serial order, let $\State_i$ be the policy state before operation $i$.
The decision is:
\[
  \Decision_i = \Policy_i(\Req_i, \vec{\View}_i(\State_i, \Req_i))
\]

The next policy state is:
\[
  \State_{i+1} =
  \begin{cases}
    \Effect_i(\State_i, \Req_i), & \Decision_i = \Allow \\
    \State_i, & \Decision_i = \Deny
  \end{cases}
\]

Thus, \pss{} couples each terminal decision to its position in the serial explanation.
An allowed operation applies its governed effect after the policy is evaluated on the preceding policy state; a denied operation leaves the policy state unchanged and produces no governed effect.

\pss{} is intentionally a property of terminal histories, not a promise that every pending approval will eventually commit.
Revalidation at resolution is a fresh policy evaluation over the state and time certified by the provider at resolution.
An intervening effect is therefore considered whenever it remains visible in the current rolling-window view, and the operation may be denied even after human approval.
Reservations and holds provide a stronger operational property by preserving the relevant capacity while an operation waits.
A deployment may also expire old pending requests and require fresh admission; such a request-age rule is orthogonal to \pss{}.
Section~\ref{subsec:rq-pending} evaluates approval preservation separately from terminal-history correctness.

\subsection{Policy-Induced Conflicts and Sound Provider Contracts}
\label{subsec:policy-induced-conflicts}

For operation $i$, let $\Rpolicy_i$ be the canonical logical scopes read by its policy evaluation, and let $\Reffect_i$ and $\Weffect_i$ be the scopes read and written by its effect.
Conventional conflict reasoning compares the footprints of effects.
Stateful governance adds another dependency: an effect can change policy state that a concurrent authorization decision has already read.
Operations $i$ and $j$ therefore have a policy-induced conflict when $\Weffect_i$ overlaps $\Rpolicy_j$.
This conflict can arise even when their application writes are independent, so ordinary application-level conflict control may have no reason to order them.
A runtime that ignores this dependency can produce stale authorization: a policy decision may be based on a policy-state fact that a concurrent effect changes before the decided effect becomes visible.

Sound provider contracts are the link between these conflicts and the mechanisms in Section~\ref{sec:design}.
For each governed operation $i$, the selected mechanism acquires, validates, or reserves a set of logical scopes $\Scopes_i$.
Provider contracts are sound when those scopes cover all policy-state dependencies of the operation: the policy read set $\Rpolicy_i$, the effect read set $\Reffect_i$, and the effect write set $\Weffect_i$.
Equivalently, $\Rpolicy_i \cup \Reffect_i \cup \Weffect_i \subseteq \Scopes_i$.
Thus, every policy-state conflict maps to at least one shared scope; extra scopes are safe but may reduce concurrency.
Validation means that before a terminal decision is recorded, and before an allowed effect commits, the mechanism verifies that the relevant scoped policy and effect state is unchanged from the state used for evaluation, has been reserved for the operation, or has been continuously protected by scoped enforcement.
If validation fails, the operation must re-evaluate, deny, or remain pending rather than commit on the stale basis.

\begin{theorem}[Policy-State Serializability by Scoped Enforcement]
\label{thm:pss-scoped-enforcement}
If deployed policies are well typed and bounded, provider contracts are sound, and the selected enforcement mechanisms serialize, reserve, or validate overlapping logical scopes before recording a terminal decision and before committing any allowed effect, then every terminal history produced by the runtime satisfies \pss{}.
\end{theorem}

Theorem~\ref{thm:pss-scoped-enforcement} states the obligation that the design must realize: identify the policy-state dependencies induced by certified policy-state views and provider effects, then protect overlapping scopes before a terminal decision and any allowed effect become visible.
The proof appears in Appendix~\ref{app:correctness-proof}.

\section{\sys{} Design}
\label{sec:design}

Section~\ref{sec:model} states the obligation that an enforcing runtime must realize: identify the policy-state dependencies used by a decision, protect the corresponding logical scopes, and commit the policy-state effect only when that decision remains valid.
\sys{} implements this obligation as a governance layer between policy authors, policy-state providers, and concurrent agentic frameworks.
It does not attempt to own the entire agent workflow.
Instead, it owns the governance path: request normalization, dependency and scope resolution, scoped enforcement, policy evaluation, policy-state commit, and governance records.

\subsection{Runtime Boundary and Roles}
\label{subsec:runtime-boundary}

\begin{figure}[t]
  \centering
  \includegraphics[width=0.78\linewidth]{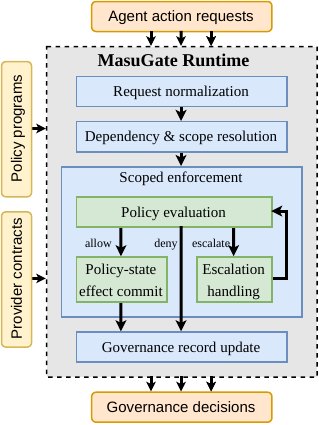}
  \caption{\sys{} runtime path for a governed action request, including escalation handling before a terminal decision.}
  \label{fig:masugate-runtime}
\end{figure}

Figure~\ref{fig:masugate-runtime} focuses on the runtime path for a single governed operation.
The broader deployment boundary is the one introduced in Figure~\ref{fig:governance-schema}: policy authors supply policy programs, policy-state providers supply contracts for certified policy-state views and governed effects, and concurrent agentic frameworks submit action requests and receive governed results.
Applications and external services remain outside the \sys{} runtime boundary.

Within that boundary, an action request is first normalized into a principal, action, arguments, and idempotency key.
The runtime then combines the compiled policy program with provider contracts to resolve dependencies and logical scopes.
Scoped enforcement begins after scope resolution and remains active while the policy is evaluated and, for an allowed operation, while the policy-state effect is committed.
Denied operations produce governance records without committing the effect.
Escalated operations enter the escalation-handling path shown in Figure~\ref{fig:masugate-runtime}; after waiting, reservation consumption, or revalidation, they resolve to a terminal decision.
Allowed operations commit the policy-state effect and then record the terminal result.

This separation is the main abstraction boundary.
Policy authors can change policy programs without rewriting provider-owned effect implementations.
Policy-state providers expose certified policy-state view and effect contracts without exposing arbitrary storage internals to policy authors.
The agentic framework sees a governed result, but it does not receive a reusable allow token detached from the policy-state commit.
This boundary also defines the threat model: agents and agent frameworks are untrusted callers, while the \sys{} runtime and policy-state providers are trusted to implement certified views, scopes, effects, reservations, and idempotency.
The \pss{} guarantee requires complete mediation: every governed effect and every mutation of policy state visible to certified views must pass through a \sys{}-mediated provider path.
Direct database writes, administrative scripts, or alternate tools that mutate the same policy state are outside the guarantee unless the provider synchronizes them with the same contracts.

\subsection{Policy Programs and Provider Contracts}
\label{subsec:contracts}

\sys{} relies on contracts rather than arbitrary policy callbacks.
A certified policy-state view contract declares a typed view interface, a trusted resolver that returns the view value, and a scope resolver that maps view arguments to a finite set of logical policy-state scopes.
An effect contract declares the action schema, the required consistency guarantee, a policy-state footprint resolver, and the trusted executor for the governed policy-state effect.
For reservation-based enforcement, view contracts also identify whether a view represents available capacity, consumed capacity, or state that is guarded again at commit.
These definitions match the prototype interface: views and effects are registered in a contract registry, and the coordinator uses those registrations to evaluate policies and commit effects.

Certification is operational rather than magical: the provider declares the view's type, boundedness, resolver, scope-set mapping, and reservation role, and the runtime admits only policy calls to registered certified views.
The compiler can check that policies are well typed and bounded over those declarations, and the coordinator can check that every selected view and effect has a scope resolver.
The runtime still trusts the provider implementation to return the promised value and to name scopes soundly; that trusted boundary is the price of exposing useful state without letting policies call arbitrary storage code.

Logical scopes are synchronization names over policy state.
They need not be physical database rows or locks.
For example, a budget view may map a team and time window to one or more scopes, often a single scope such as \texttt{team-budget:research}, while the transfer effect declares that committing a transfer writes the same budget scope.
The runtime uses scopes to decide which operations must be coordinated; the provider code decides how view values and effect footprints are computed.

Figure~\ref{fig:provider-contract-example} sketches the contract shape for the running transfer policy.
The key point is that the budget views and the transfer effect declare a shared logical budget scope, so the runtime can coordinate the policy read with the effect commit.
The provider also declares idempotency and the supported enforcement mode; these are trusted provider obligations, not policy-program logic.

\begin{figure}[t]
\centering
\begin{minipage}{0.96\linewidth}
\begin{lstlisting}[style=masugatepolicy,
  morekeywords={view,scope,scopes,role,effect,reads,writes,idempotency,enforcement,transaction,reservation}]
view budget.spent(team, window)
  scopes { team-budget:{team} }
  role consumed_capacity

view budget.limit(team)
  scopes { team-budget:{team} }
  role capacity_limit

effect transfer(sender, receiver, amount, team)
  reads  team-budget:{team}
  writes team-budget:{team}
  idempotency request_id
  enforcement transaction | reservation
\end{lstlisting}
\end{minipage}
\caption{Example provider contract for the running budget-transfer policy.}
\label{fig:provider-contract-example}
\end{figure}

The trusted obligations break down into view soundness, scope soundness, effect soundness, reservation soundness, and idempotency and recovery soundness.
The view must return the promised policy-state fact, the scope resolvers must name every logical dependency, the effect must update policy state consistently with the visible outcome, reservations must prevent double consumption, and retries must not duplicate effects.
Reservation soundness is escrow-style~\cite{oneil1986escrow}: creating a reservation atomically removes capacity from the amount available to other operations, consuming it applies the reserved effect at most once, and canceling it returns the capacity.

Intuitively, a provider contract is sound when it gives \sys{} every scope name needed to protect the relevant policy state.
If a policy may read a policy-state fact, or an effect may read or update that fact, the corresponding view or effect contract must map the request to a set of scopes that covers it.
Extra scopes are safe but may reduce concurrency.
If a required resolver is absent, \sys{} rejects the binding during admission; if a resolver is present but omits a needed scope, the runtime can protect only the declared scopes, so the deployment violates the contract and the guarantee in Theorem~\ref{thm:pss-scoped-enforcement} no longer applies.

\subsection{Dependency and Scope Resolution}
\label{subsec:policy-dsl}

Dependency resolution has both compile-time and request-time parts.
At compile time, the policy compiler validates the policy and records the finite set of certified policy-state view calls that may be evaluated.
At request time, \sys{} evaluates request-dependent arguments, such as \texttt{principal.team}, and invokes the corresponding provider scope resolvers.
The result is a concrete set of policy-read scopes for this operation.

The effect contract supplies the other half of the dependency set.
For the requested action, its footprint resolver maps the action request to the policy-state scopes that the effect may read or write.
The runtime combines policy-read scopes, effect scopes, and an idempotency scope into the operation's coordination set.
The current prototype implements this flow directly: the policy runtime computes dependency scopes from compiled host calls, the effect contract computes a policy-state footprint, and the coordinator protects their union.

The policy interface is intentionally restricted to make this extraction possible.
Policies are pure, bounded programs over the request, principal attributes, action arguments, and registered certified policy-state views.
They cannot call arbitrary provider code, mutate state, or loop over unbounded data.
Figure~\ref{fig:policy-dsl-core} gives the core policy-language fragment; Appendix~\ref{app:policy-language-details} gives the full DSL syntax and static checks used by the prototype.

\begin{figure}[t]
  \centering
  \begingroup
  \setlength{\fboxsep}{2.5pt}
  \setlength{\fboxrule}{0.3pt}
  \fbox{%
    \begin{minipage}{0.88\linewidth}
    {\scriptsize
    \[
    \begin{array}{rcl}
    p &::=& \mathsf{policy}\ n\ \mathsf{on}\ a\ \{\overline{r}\; d\}\;\\
    d &::=& \mathsf{allow}\ \mathsf{otherwise}\\
    r &::=& \mathsf{deny}\ \rho\ \mathsf{when}\ e
        \mid \mathsf{escalate}\ \rho\ \mathsf{when}\ e\\
    e &::=& c \mid \Args.x \mid \Prin.x \mid Q(e_1,\ldots,e_k)\\
      &\mid& e_1\ \mathit{op}\ e_2 \mid \neg e\\
    Q &\in& \mathcal{V}_{\mathsf{cert}}
    \end{array}
    \]
    }
    \end{minipage}%
  }
  \endgroup
  \caption{Core policy-language fragment. Certified view symbols $Q \in \mathcal{V}_{\mathsf{cert}}$ are registered by provider contracts with type and scope resolvers; all expressions are bounded and side-effect free.}
  \label{fig:policy-dsl-core}
\end{figure}

\subsection{Scoped Enforcement}
\label{subsec:execution-protocol}

Scoped enforcement is the protected interval in Figure~\ref{fig:masugate-runtime}.
After resolving scopes, \sys{} enters an enforcement context for those scopes.
Within that context, it evaluates the policy over certified policy-state views and then handles the resulting decision.
A denied operation records a denial and produces no policy-state effect.
An escalated operation enters the pending path, optionally with a reservation, and later re-enters enforcement at resolution.
An allowed operation invokes the trusted effect executor and commits the policy-state effect before returning the allowed result to the agentic framework.

The prototype realizes scoped enforcement with two layers.
The coordinator first acquires local keyed locks over the resolved scopes to serialize in-process conflicts.
For backends that expose database-level scoped locks, such as the PostgreSQL ledger, the coordinator also asks the resource to acquire scoped locks inside the resource session.
The same session is used for policy evaluation, effect execution, reservation updates, idempotency checks, and governance record writes.

\sys{} supports different enforcement modes behind the same contract.
Transaction mode evaluates the policy and commits the allowed effect inside one provider-owned transaction protected by the resolved scopes.
Reservation mode creates or consumes durable capacity records for long-running approvals or contended resources; startup admission rejects bindings whose policies do not use provider-declared reservation-safe view patterns.
Both modes use scoped locking for overlapping logical scopes and durable idempotency records for retries.

\subsection{Governance Records, Escalation, and Evolution}
\label{subsec:long-running-governance}

Governance records are durable runtime state written by the enforcement protocol.
They link the request, policy and rule identifiers, certified policy-state view reads, decision, and policy-state effect that the runtime enforced.
Together, these records are the audit counterpart of scoped enforcement: they describe the same operation whose scopes and effect were protected.

Escalation is treated as a split-phase governed operation.
The initial policy evaluation records a pending operation rather than a terminal allow.
In reservation mode, \sys{} also reserves the relevant capacity; in transaction mode, the pending operation must revalidate policy state before committing.
At resolution, the operation consumes the reservation or revalidates policy state before committing the effect, and only then becomes terminal in the history used by \pss{}.

The same boundary supports policy evolution.
Adding a new guard should usually require changing a policy and, when necessary, registering a new certified policy-state view, not rewriting every provider-owned effect.
Because governance records preserve structured decision evidence, \sys{} can keep policies first class while still producing audit records for review.

\subsection{Prototype Realization}
\label{subsec:prototype-realization}
We implemented \sys{} as a Python prototype with a bounded policy frontend, a contract registry, a policy runtime, and a governed-operation coordinator.
The PostgreSQL backend stores policy state, pending operations, reservations, idempotency records, and audit records in tables, and protects logical scopes with transactions and transaction-scoped advisory locks~\cite{postgresExplicitLocking2026}.
A SQLite backend supports local smoke tests, and a thin MAF-shaped adapter~\cite{microsoftAgentFramework2026} maps framework function calls into \sys{} action requests.
Section~\ref{sec:evaluation} evaluates the PostgreSQL-backed prototype.

\section{Evaluation}
\label{sec:evaluation}

The evaluation is intentionally controlled: each workload isolates one part of the correctness obligation rather than attempting to mimic an entire deployment trace.
The five research questions should be read in that order.
\textbf{RQ1} tests the core safety failure, stale authorization.
\textbf{RQ2} measures the cost of preserving that safety for synchronous operations.
\textbf{RQ3} separates terminal-history correctness from the stronger need to preserve a delayed approval's policy-state basis.
\textbf{RQ4} tests whether the boundary keeps policy changes mostly outside trusted provider code.
\textbf{RQ5} checks whether the same properties hold in a multi-policy agentic workflow.
We study these questions with PostgreSQL-backed controlled workloads and a scripted, LLM-free agentic procurement benchmark.

\subsection{Common Experimental Setup}
\label{subsec:evaluation-setup}

\textbf{Backend and run configuration.}
The reported results use the PostgreSQL backend of the \sys{} research prototype described in Section~\ref{subsec:prototype-realization}.
It stores policy state in scope-indexed tables and enforces logical scopes with PostgreSQL transactions and transaction-scoped advisory locks~\cite{ports2012ssi,postgresExplicitLocking2026}.
Unless otherwise stated, numeric aggregates are means over five random seeds; stale-allow columns report the maximum over seeds, and \pss{} columns require all seeds to satisfy \pss{}.
The machine has an Intel Core Ultra 7 155H CPU with 22 logical CPUs and runs under WSL2.
The external baseline environment uses Cedar CLI 4.11.1, Agent Governance Toolkit 4.1.0, and Omnigent 0.4.0~\cite{cedar2024,agt2026,omnigent2026}.

\smallskip
\noindent\textbf{Baselines and modes.}
We group baselines by role.
Request-local baselines test policy-engine integrations in which mutable policy state is supplied by the caller and the database effect follows the returned decision.
Naive check-then-act (\textsc{Naive}) evaluates the stateful policy directly without coordination, while RQ1 uses Cedar as a representative external policy engine in this configuration.
RQ5 separately evaluates AGT and Omnigent using their native cost-governance mechanisms.
Correctness-preserving baselines test the cost of safe enforcement.
Global serialization (\textsc{Global}) protects the check-effect window with one coarse critical section.
Manual application transactions (\textsc{Manual tx}) embed one fixed policy check inside trusted provider code using the same logical scopes as the governed effect.
\sys{} modes keep policies first-class and rely on certified policy-state views and provider-declared effect footprints.
\systx{} enforces the check-effect window with scoped transactional coordination.
\sysres{} adds provider-defined reservations for consumable policy state.
For long-running approval, \syshold{} creates durable holds on affected logical scopes while approval is outstanding.

\smallskip
\noindent\textbf{Metrics.}
Correctness metrics track stale allows, terminal outcomes, and \pss{} compatibility.
Performance metrics track throughput, latency, retries, and coordination wait.
Pending-operation metrics track approval invalidation, unrelated progress, scope-hold waits, and reservation behavior.
Policy-evolution metrics track trusted provider-code edits and audit-read coverage.
Agentic-workflow metrics track valid committed workflows, stale allows, stale approvals, final policy-state violations, and throughput.

\subsection{RQ1: Does \sys{} Prevent Stale Authorization?}
\label{subsec:rq-correctness}

\textbf{Design.}
The minimal-conflict workload is the smallest stale-authorization anomaly.
It initializes a team close to its daily budget and runs two concurrent transfers that are each individually allowed if checked before either effect commits, but only one transfer can be valid in any serial policy-state history.
The two transfers debit different accounts, so their application writes are disjoint; they conflict only because both consume the same team-budget policy-state scope.
A correct mode should commit one transfer, deny one transfer, report no stale allows, and satisfy \pss{}.
We also run a many-client full-conflict workload in which 256 operations share the same policy-state scope; a correct mode should admit only the operations that fit within the budget.

\smallskip
\noindent\textbf{Results.}
Table~\ref{tbl:correctness-results} reports the minimal-conflict and full-conflict checks.
In the minimal conflict, \textsc{Naive} and \textsc{Cedar} commit both racing transfers, while every correctness-preserving mode commits one transfer and denies the other.
In the full-conflict workload, \textsc{Naive} and \textsc{Cedar} produce 30--31 stale allows and commit 79.4--80.8 transfers although the budget admits only 50.
Global serialization, manual fixed-policy transactions, and both \sys{} modes commit exactly 50 transfers, deny the rest, report zero stale allows, and satisfy \pss{}.
Thus RQ1 answers yes for \sys{}: scoped policy-state coordination removes stale authorization in both the minimal race and the multiprocess full-conflict workload.

\begin{table}
  \caption{Correctness under minimal and full-conflict policy-state races. C/D is committed/denied operations; full-conflict C/D values are means over five random seeds, and Stale is the maximum stale-allow count over seeds.}
  \label{tbl:correctness-results}
  \footnotesize
  \begin{tabular}{@{}lcccccc@{}}
    \toprule
    & \multicolumn{3}{c}{Minimal conflict} & \multicolumn{3}{c}{Full conflict} \\
    \cmidrule(lr){2-4}\cmidrule(l){5-7}
    Mode & C/D & Stale & \pss{} & C/D & Stale & \pss{} \\
    \midrule
    \textsc{Naive} & 2/0 & 1 & \(\times\) & 79.4/176.6 & 30 & \(\times\) \\
    \textsc{Cedar}~\cite{cedar2024} & 2/0 & 1 & \(\times\) & 80.8/175.2 & 31 & \(\times\) \\
    \textsc{Global} & 1/1 & 0 & \(\surd\) & 50/206 & 0 & \(\surd\) \\
    \textsc{Manual tx} & 1/1 & 0 & \(\surd\) & 50/206 & 0 & \(\surd\) \\
    \systx{} & 1/1 & 0 & \(\surd\) & 50/206 & 0 & \(\surd\) \\
    \sysres{} & 1/1 & 0 & \(\surd\) & 50/206 & 0 & \(\surd\) \\
    \bottomrule
  \end{tabular}
\end{table}

\subsection{RQ2: What Is the Cost of First-Class Governance?}
\label{subsec:rq-overhead}

\textbf{Design.}
RQ2 isolates synchronous governed operations, with no human approval and no policy exhaustion.
Each cell runs 512 transfers with 32 clients over 16 logical scopes, so two random operations are disjoint with probability 0.9375.
We vary governed-operation service time from 0 to 10 ms to model work that must occur inside the protected check-effect window.
All compared modes are correctness preserving: \textsc{Manual tx} is the specialized fixed-policy transaction baseline, \textsc{Global} is the generic serialization baseline, and \systx{} is the main policy-first \sys{} path.
We also include \sysres{} to measure the cost of reservation bookkeeping when no pending window needs it.

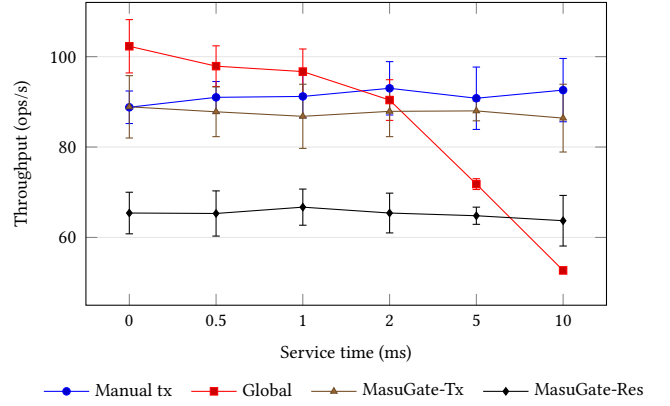
\begin{figure}
  \centering
  \begin{tikzpicture}
    \begin{axis}[
      width=\columnwidth,
      height=0.66\columnwidth,
      xlabel={Service time (ms)},
      ylabel={Throughput (ops/s)},
      symbolic x coords={0,0.5,1,2,5,10},
      xtick=data,
      ymin=45, ymax=112,
      ymajorgrids=true,
      grid style={black!10},
      legend columns=4,
      legend style={font=\scriptsize, draw=none, at={(0.5,-0.22)}, anchor=north, /tikz/every even column/.append style={column sep=0.5em}},
      tick label style={font=\scriptsize},
      label style={font=\scriptsize},
      mark size=1.4pt,
    ]
      \addplot+[mark=*, error bars/.cd, y dir=both, y explicit] coordinates {(0,88.8) +- (0,3.6) (0.5,91.0) +- (0,3.5) (1,91.2) +- (0,4.7) (2,93.0) +- (0,5.9) (5,90.8) +- (0,6.9) (10,92.6) +- (0,7.0)};
      \addlegendentry{Manual tx}
      \addplot+[mark=square*, error bars/.cd, y dir=both, y explicit] coordinates {(0,102.3) +- (0,5.9) (0.5,97.9) +- (0,4.5) (1,96.7) +- (0,5.0) (2,90.4) +- (0,4.5) (5,71.8) +- (0,1.2) (10,52.7) +- (0,0.7)};
      \addlegendentry{Global}
      \addplot+[mark=triangle*, error bars/.cd, y dir=both, y explicit] coordinates {(0,88.9) +- (0,6.9) (0.5,87.8) +- (0,5.5) (1,86.8) +- (0,7.1) (2,87.9) +- (0,5.6) (5,88.0) +- (0,2.2) (10,86.4) +- (0,7.5)};
      \addlegendentry{MasuGate-Tx}
      \addplot+[mark=diamond*, error bars/.cd, y dir=both, y explicit] coordinates {(0,65.4) +- (0,4.6) (0.5,65.3) +- (0,5.0) (1,66.7) +- (0,4.0) (2,65.4) +- (0,4.4) (5,64.8) +- (0,1.9) (10,63.7) +- (0,5.6)};
      \addlegendentry{MasuGate-Res}
    \end{axis}
  \end{tikzpicture}
  \caption{RQ2 service-time sweep at 32 clients and 16 logical scopes. Values are means over five random seeds; error bars show one standard deviation. Every row commits all 512 operations with zero violations.}
  \label{fig:rq2-delay-sweep}
\end{figure}

\smallskip
\noindent\textbf{Results.}
All compared modes commit all 512 operations and report zero violations, denies, pending operations, or aborts.
Figure~\ref{fig:rq2-delay-sweep} shows the throughput sweep.
At 0 ms service time, \systx{} reaches 88.9 ops/s, matching \textsc{Manual tx} at 88.8 ops/s and reaching 0.87x \textsc{Global}.
As service time grows, global serialization pays for the protected delay directly: at 10 ms, \textsc{Global} falls to 52.7 ops/s.
Because \textsc{Manual tx} and \systx{} protect only the operation's declared scopes, much of the added service time can proceed in parallel across the 16 logical scopes.
At 10 ms, \systx{} reaches 86.4 ops/s, or 0.93x \textsc{Manual tx} and 1.64x \textsc{Global}.
The latency and wait columns in the artifact remain useful diagnostics, but we do not use p95 latency as an RQ2 headline: in repeated artifact runs, per-seed tail values were sensitive to seed-local benchmark-runner stalls that moved between modes, while completion counts, violation counts, and throughput were stable.
The absolute throughput reflects prototype overheads common to these paths, including Python request orchestration, PostgreSQL session work, scoped lock acquisition, and governance-record writes.
\sysres{} is slower than \systx{} throughout this synchronous workload because it pays reservation bookkeeping without a pending approval to preserve, but at 10 ms it still reaches 1.21x global throughput.

The overall result is that \sys{} pays measurable overhead when work is very short, but scoped policy-first enforcement avoids the coarse serialization cost that dominates global locking as governed service time increases.

\subsection{RQ3: Does \sys{} Support Long-Running Governance?}
\label{subsec:rq-pending}

\textbf{Design.}
The pending-approval experiments test whether a runtime can preserve an approval's policy-state basis while still making progress on independent work.
The failure mode is stale approval, the split-phase version of stale authorization.
In a procurement workflow, for example, a human manager may approve a purchase based on capacity available when the request enters review, but another workflow may consume that capacity before the approved effect commits.
Treating the approval as a timeless token can therefore produce a policy-violating commit; pure revalidation avoids that commit, but can reject an already approved operation and create manual follow-up work.
The workloads below isolate this tradeoff.
The single-approval workload has one transfer pause for approval, 16 unrelated transfers on disjoint policy-state scopes, and one same-scope competitor.
The scope-scaling workload runs 64 clients across 1 to 64 teams, marks 10\% of transfers as pending, and uses a 1s approval delay.
The desired behavior is concrete: unrelated transfers should make progress during the approval wait, same-scope competitors should not invalidate pending approvals, and approved transfers should commit once approval arrives.

The modes form a coordination ladder.
Global hold keeps one global critical section across the approval wait; global revalidation releases it and checks again at approval resolution.
\systx{} revalidates pending approvals using provider-certified state and time at resolution.
\syshold{} creates durable holds on the pending transfer's logical scopes, allowing disjoint scopes to proceed while same-scope competitors wait.
\sysres{} uses provider-defined reservations, so a same-scope competitor can be denied immediately when it conflicts with reserved capacity.

\begin{figure}
  \centering
  \includegraphics[width=\columnwidth]{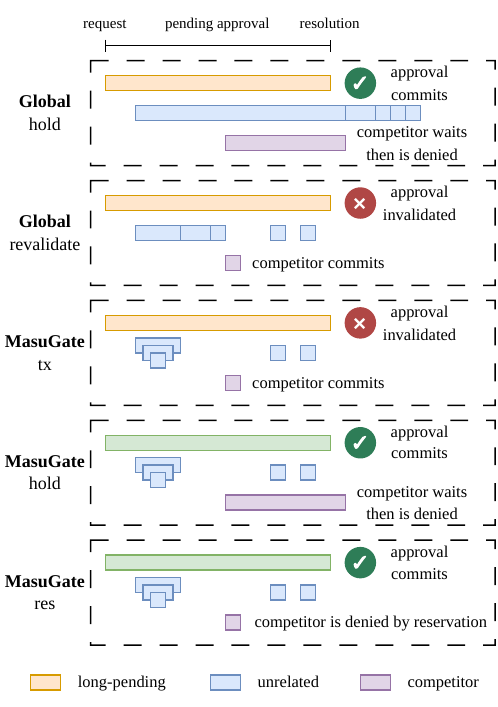}
  \caption{Pending-approval benchmark. One transfer waits for approval while unrelated transfers and a same-scope competitor run.}
  \label{fig:pending-approval}
\end{figure}

\smallskip
\noindent\textbf{Results.}
Figure~\ref{fig:pending-approval} illustrates the single-approval benchmark.
Global hold preserves the approval, but no unrelated transfer commits before approval resolution and unrelated p95 latency rises to 1080.8 ms.
Global revalidation and \systx{} allow all 16 unrelated transfers to commit before the approval resolves, but the same-scope competitor commits first and invalidates the pending approval.
\syshold{} and \sysres{} both preserve the approval while allowing all 16 unrelated transfers to commit before resolution.
The difference is the same-scope path: \syshold{} makes the competitor wait for the approval and then denies it, with 1021.8 ms p95 scope-hold wait, while \sysres{} denies the competitor before resolution using the reservation.

Table~\ref{tbl:pending-scope-scale} separates the two properties the scope-scaling workload is meant to test.
The first numeric column reports approval invalidation rate: the fraction of pending approvals whose policy-state basis is lost before resolution.
Pending-commit rate is omitted because the two rates sum to 100\% in this workload.
The remaining columns report fixed-window unrelated progress: unrelated transfers committed during the configured 1s approval window divided by unrelated transfers submitted, across all nontrivial team counts in the 2-to-64-team sweep.
This avoids crediting a mode for unrelated work that commits only because approval resolution itself was delayed behind runtime coordination.
Global hold preserves approvals but makes no unrelated progress during the wait.
Global revalidation and \systx{} permit same-scope competitors to commit before approval resolution, so every pending approval loses its basis; their higher fixed-window progress is therefore not approval-preserving progress.
\syshold{} and \sysres{} are the only modes that preserve every pending approval, but this guarantee has an engineering cost under high contention.
Their lower fixed-window progress reflects prototype queueing and PostgreSQL coordination overhead while many approval-bearing and same-scope competing operations are outstanding.
Reservations reduce this cost for consumable state by denying reserved-capacity conflicts immediately: at 64 teams, \sysres{} preserves approvals with 12.4\% fixed-window progress, compared with 7.5\% for \syshold{}.

\begin{table}
  \caption{Scope-scaling pending workload. Entries are percentages averaged over five random seeds. Progress is unrelated work committed during the fixed 1s approval window divided by unrelated work submitted. Bold nonzero invalidation entries mark modes that fail to preserve pending approvals.}
  \label{tbl:pending-scope-scale}
  \scriptsize
  \setlength{\tabcolsep}{1.2pt}
  \begin{tabular}{@{}lccccccc@{}}
    \toprule
    \multirow{2}{*}{Mode} & \multirow{2}{*}{\shortstack{Approval\\invalidation\\rate}} & \multicolumn{6}{c}{Unrelated progress by team count} \\
    \cmidrule(lr){3-8}
    & & 2 & 4 & 8 & 16 & 32 & 64 \\
    \midrule
    Global hold & 0\% & 0\% & 0\% & 0\% & 0\% & 0\% & 0\% \\
    Global reval. & \textbf{100\%} & 100\% & 100\% & 100\% & 89.5\% & 42.7\% & 13.9\% \\
    \systx{} & \textbf{100\%} & 100\% & 100\% & 100\% & 75.0\% & 36.1\% & 18.0\% \\
    \syshold{} & 0\% & 100\% & 100\% & 100\% & 39.1\% & 15.5\% & 7.5\% \\
    \sysres{} & 0\% & 100\% & 100\% & 100\% & 56.2\% & 26.8\% & 12.4\% \\
    \bottomrule
  \end{tabular}
\end{table}

The takeaway is that pending approval needs more than transaction-mode correctness.
Revalidation can avoid stale commits and therefore preserve \pss{}, but it may invalidate the human approval because the policy-state basis has changed.
Scoped hold is a correct fallback, but its wait-based treatment of same-scope competitors can reduce short-window progress at scale.
Reservations are preferable for consumable state: they preserve the pending approval while denying conflicting same-scope work without a long wait.

\subsection{RQ4: Does \sys{} Preserve Policy Modularity?}
\label{subsec:rq-evolution}

\textbf{Design.}
The policy-evolution workload tests whether the \sys{} boundary keeps policy changes out of trusted provider code while preserving executable policy behavior.
Starting from the team-budget policy, we apply five representative changes: a per-agent rolling limit, an approval-state guard, a risk-score guard, a 24-hour to 7-day window change, and a second governed action.
For each variant, the harness parses and compiles the \sys{} policy, runs targeted policy tests over controlled certified-view fixtures, records audit-read capture, and counts changed lines in policy code and trusted provider code.
The comparison point is manual fixed-policy enforcement, where the policy logic is embedded in trusted provider code.
This is a source-evolution and testability experiment rather than a randomized runtime benchmark.
Appendix~\ref{app:policy-evolution-fixtures} lists the policy programs and expert baseline programs used to compute these source deltas.

\smallskip
\noindent\textbf{Results.}
All six \sys{} variants compile, all 23 policy tests pass, and audit reads are captured automatically for every variant.
Across the five evolution tasks, \sys{} changes 18 policy lines but only 4 trusted provider-code lines, while manual fixed-policy enforcement changes 22 trusted provider-code lines.
Table~\ref{tbl:policy-evolution-results} puts provider-contract additions, policy-text churn, and trusted-code churn side by side. In the table, $\Delta$ denotes changed source lines relative to the base team-budget fixture; the main signal is that policy changes remain mostly outside trusted provider code.
The line counts are a proxy for boundary movement, not a universal productivity metric: the stronger evidence is that each new policy-state input is introduced as a certified view and then appears automatically in the policy's audit-read trace.
For example, the risk-score variant adds a \texttt{risk.score} view contract and a policy rule, while the transfer effect executor remains unchanged and the resulting audit trace records the risk read alongside the budget reads.

\begin{table}
  \caption{Policy-code and trusted-code changes under policy evolution. Contracts added are provider-facing certified-view or action contracts; $\Delta$ columns report changed source lines relative to the base team-budget fixture.}
  \label{tbl:policy-evolution-results}
  \scriptsize
  \setlength{\tabcolsep}{2.2pt}
  \begin{tabular}{@{}lrrrr@{}}
    \toprule
    Change & \shortstack{Contracts\\added} & \shortstack{MasuGate\\policy $\Delta$} & \shortstack{MasuGate\\trusted $\Delta$} & \shortstack{Manual\\trusted $\Delta$} \\
    \midrule
    Per-agent limit & 1 & 2 & 1 & 3 \\
    Approval guard & 1 & 2 & 1 & 3 \\
    Risk guard & 1 & 2 & 1 & 3 \\
    7-day window & 0 & 4 & 0 & 4 \\
    Second action & 0 & 8 & 1 & 9 \\
    \midrule
    Total & 3 & 18 & 4 & 22 \\
    \bottomrule
  \end{tabular}
\end{table}

Adding new policy-state inputs, such as per-agent history, approval count, or risk score, requires \sys{} to register a small provider contract while leaving the transfer effect executor unchanged.
For the first four variants, the trusted effect path is stable; the changes are policy text, certified-view registrations, or fixture metadata.
Changing the time window is a pure policy edit in \sys{} but a trusted-code edit in the manual baseline, because the latter embeds the policy predicate inside the provider implementation.
The second action is the largest semantic change because it adds a new governed action and rules; in \sys{} most of that change remains policy text, while the trusted provider change is one line.
Thus RQ4 does not claim policy evolution is free; it shows that \sys{} makes the evolution testable and auditable while sharply reducing trusted-code churn.

\subsection{RQ5: Does \sys{} Preserve Governance in Agentic Workflows?}
\label{subsec:agentic-workflow}

\textbf{Design.}
RQ5 moves from single-policy microbenchmarks to a scripted, LLM-free procurement workflow inspired by tool-agent evaluation settings~\cite{taubench2024,agentdojo2024,toolemu2023}.
Each workflow carries task metadata through an explore, refine, govern, and commit shape: it chooses a preferred item, may fall back when inventory is unavailable, checks daily agent-cost, team-budget, inventory, and approval policies, waits for simulated reasoning or human review, and then commits cost, order, budget, and inventory effects.
The workload is still controlled rather than a live LLM deployment, which lets us fix the conflict rate and directly measure stale authorization.
The reported run uses 256 workflows, 32 clients, 8 teams, 16 items, five seeds, 50\% hot teams and items, 20\% approval-triggering workflows, a 50\,ms reasoning delay, and a 1\,s approval delay.
Each workflow costs \$0.10 against a \$1.00 per-agent daily limit.
The hot user begins at \$0.80, leaving capacity for two additional workflows, while the hot team begins at 9,000 of its 10,000-unit budget, leaving capacity for one order.
The unequal remaining capacities prevent enforcement of daily cost from masking stale team-budget decisions.

The comparison focuses on agent-governance baselines.
\textsc{AGT} uses the AGT policy engine for team-budget and inventory decisions and its native \texttt{CostGuard} for per-agent daily cost.
One shared \texttt{CostGuard} instance per run atomically checks and charges its in-process budget state before the reasoning or approval delay; the admitted workflow later applies its PostgreSQL effects.
\textsc{Omnigent} uses its native per-user daily-cost policy~\cite{omnigent2026} together with Omnigent-style \textsc{ALLOW}, \textsc{DENY}, and \textsc{ASK} verdicts.
In both configurations, team-budget, inventory, and approval decisions precede the corresponding PostgreSQL effects.
\systx{} rechecks and commits inside scoped PostgreSQL coordination, while \sysres{} reserves daily cost, team budget, and inventory capacity before the delayed work or approval.

\smallskip
\noindent\textbf{Results.}
Table~\ref{tbl:rq5-agentic-workflow} reports the aggregate result.
\textsc{AGT} commits 71.6 workflows on average, of which 59.0 are valid.
Its native \texttt{CostGuard} eliminates daily-cost violations in all five seeds, but stale policy-engine decisions leave one team and one inventory item invalid in every run.
\textsc{Omnigent} commits 107.6 workflows, of which 66.2 are valid, and produces final daily-cost, team-budget, and inventory violations.
Omnigent's native cost policy therefore improves the policy vocabulary, but the run still exposes the missing decision-effect binding.
\systx{} and \sysres{} each commit 70.6 valid workflows with no stale decisions or final policy-state violations.
The approval columns show that AGT preserves 11.6 approval bases on average, Omnigent preserves 1.6, and both \sys{} modes lose none of the approvals they create; \sysres{} preserves 13.8 through reservations.

\begin{table}
  \centering
  \caption{Agentic procurement workflow benchmark. Values are means over five seeds except Viol., which reports maxima. Stale auth. counts committed workflows whose policy-state basis became invalid; Appr., Stale appr., and Appr. pres. count approval-created workflows, lost approval bases, and preserved approval bases. Viol. reports final invalid daily-cost users, teams, and inventory items.}
  \label{tbl:rq5-agentic-workflow}
  \scriptsize
  \setlength{\tabcolsep}{4pt}
  \begin{tabular}{@{}lrrrrrrcc@{}}
    \toprule
    Mode & Commit & \shortstack{Valid\\commit} & \shortstack{Stale\\auth.} & Appr. & \shortstack{Stale\\appr.} & \shortstack{Appr.\\pres.} & Viol. & \pss{} \\
    \midrule
    \textsc{AGT}~\cite{agt2026} & 71.6 & 59.0 & 12.6 & 14.2 & 2.6 & 11.6 & 0/1/1 & $\times$ \\
    \textsc{Omnigent}~\cite{omnigent2026} & 107.6 & 66.2 & 41.4 & 20.8 & 19.2 & 1.6 & 8/8/1 & $\times$ \\
    \systx{} & 70.6 & 70.6 & 0 & 4.4 & 0 & 4.4 & 0/0/0 & $\surd$ \\
    \sysres{} & 70.6 & 70.6 & 0 & 13.8 & 0 & 13.8 & 0/0/0 & $\surd$ \\
    \bottomrule
  \end{tabular}
\end{table}

RQ5 shows that native cost controls can protect their own state while external team-budget and inventory decisions remain stale.
\sys{} preserves all three invariants by extending protection through effect commit.

\section{Discussion and Limitations}
\label{sec:discussion}

\paragraph{Evaluation Scope}
\label{subsec:threats}
The workloads are controlled benchmarks rather than deployment traces.
This choice isolates stale authorization, policy-state sharing, approval delay, and provider-code changes; the RQ5 procurement benchmark adds multi-step task traces, policy branches, approval delays, AGT and Omnigent baselines, and shared policy state, but remains simulated rather than an open-ended agent deployment.

The prototype evaluates one concrete policy-state provider and source-level policy-evolution fixtures.
Other consumable or history-dependent policies should fit the same contract model, but production use would need additional certified views, provider adapters, and deployment hardening.
The artifact emits raw events, per-run summaries, paper-facing tables, and run metadata for reproducibility.

\paragraph{Provider Support}
\label{subsec:provider-support}
\sys{} cannot create consistency guarantees that the policy-state provider cannot enforce.
The provider must implement certified policy-state views, governed effects, and any reservation or hold mechanism needed by the selected enforcement strategy.
If a policy requires \pss{} but the provider offers only best-effort reads or unvalidated external effects, deployment should fail or the guarantee must be explicitly weakened.

This is the main trusted boundary in \sys{}.
Policies remain first-class and auditable, but provider contracts and implementations are trusted code.
The benefit is that this trusted code is shared across policies and evolves less often than policy text; the cost is that new kinds of policy-visible state require provider support.

\paragraph{Integration with Agent Frameworks}
\label{subsec:framework-integration}
\sys{} is intended to sit beside, rather than replace, an agent orchestration framework.
Frameworks such as MAF~\cite{microsoftAgentFramework2026} and LangGraph~\cite{langGraph2026} can continue to provide graph execution, task state, checkpointing, durable resumption, human review, middleware, and tool routing, while selected tool calls cross a \sys{} boundary before their governed effects commit.
These native mechanisms preserve workflow execution state but do not by themselves ensure that external mutable policy-state facts remain valid through effect commit, which is supported by \sys{} through an additional boundary.
The integration burden falls on provider adapters: they must normalize framework calls into governed requests, register the policy-state views and effect contracts for the underlying resource, and return allow, deny, or escalation outcomes to the framework.
This boundary is most useful in domains where agents act on shared resources, such as refunds, reservations, cloud capacity, access privileges, and financial transfers.

\paragraph{External Effects}
\label{subsec:external-effects}
Many external APIs cannot participate in a database transaction, and holding a database transaction open across network I/O is often undesirable.
In such cases, a better choice is a durable protected-execution protocol: protect or reserve the relevant policy state before dispatch, bind the authorization to a durable intent and external-operation identity, invoke the external operation with idempotency and recovery support, and consume, release, or quarantine the protected policy state according to the reconciled outcome.
Designing and evaluating such a protocol for \sys{} is left to future work.

\paragraph{Policy Language and Prototype Scope}
\label{subsec:policy-language-scope}
\sys{} deliberately restricts policy programs to bounded expressions over registered certified policy-state views.
This excludes arbitrary callbacks and unbounded computation in the authorization path.
The restriction is necessary so the runtime can know which policy state a decision may depend on and which coordination strategy can preserve that state.

The prototype implements the central execution paths, PostgreSQL-backed policy state, pending approval, reservations, scoped holds, and evaluation harnesses.
It does not include production deployment tooling, policy-bundle signing, or a concrete dependency on a particular agent framework API.
These omissions do not affect the \pss{} argument, but a production system would need deployment hardening and external-effect recovery.

\section{Related Work}
\label{sec:related}

\paragraph{Authorization and access control.}
Authorization policy languages and policy-as-code systems separate policy logic from application code~\cite{openPolicyAgent2026}.
Cedar~\cite{cedar2024}, Microsoft's Agent Governance Toolkit (AGT)~\cite{agt2026}, and Omnigent~\cite{omnigent2026} move governance into explicit artifacts around requests or agent actions.
AGT and Omnigent also provide stateful-policy, runtime, and workflow-oriented mechanisms beyond request evaluation.
Our evaluation isolates Cedar's request-local policy-engine integration in RQ1 and AGT's native \texttt{CostGuard} and Omnigent's native daily-cost policy in RQ5.
\sys{} differs not by merely accepting stateful inputs, but by requiring provider-certified policy-state views and effect contracts and enforcing their decision-effect semantics under \pss{}.
Database access-control work, including fine-grained authorization and predicate rewriting~\cite{rizvi2004fgac}, similarly protects what data a request may read or observe; \sys{} instead couples a policy decision to a later effect that may update the same logical policy state.

\paragraph{Transactions and coordination.}
Serializable transactions provide a direct way to couple policy reads and effects when both live behind one transactional provider~\cite{papadimitriou1979serializability}, and PostgreSQL's serializable snapshot isolation is one practical mechanism~\cite{ports2012ssi}.
Coordination-avoidance work asks when applications can avoid coordination while preserving invariants~\cite{bailis2014coordination}; Blazes identifies where distributed programs require coordination~\cite{alvaro2013blazes}; and transaction chopping decomposes transactions to improve concurrency while preserving correctness~\cite{shasha1995transaction}.
\sys{} reuses this systems lineage, but changes the abstraction boundary: the invariants are authored as governance policies over certified policy-state views, while concrete effects and coordination mechanisms remain owned by providers.

Recent agent runtimes apply transaction and concurrency-control ideas to tool-using workflows.
Atomix records reads and effects and settles them through progress-aware transactions~\cite{atomix2026}; CoAgent coordinates concurrent agents using footprint-declared, undoable tools~\cite{coagent2026}; and Cordon stages and validates tool effects using shadow state, an effect outbox, and recovery metadata~\cite{cordon2026}.
CommitGuard validates the freshness and binding of authority at the commit boundary~\cite{commitguard2026}.
\sys{} is complementary: it derives coordination requirements from policy-authored dependencies over provider-certified state, defines \pss{} as the governance correctness property, and preserves delayed approvals through scoped holds and reservations.

\paragraph{Reservations and long-running workflows.}
Escrow transactions reserve or allocate portions of shared capacity so concurrent transactions can proceed without violating aggregate invariants~\cite{oneil1986escrow}.
\sys{} uses reservations for the analogous governance problem: preserving consumable policy state, such as budget or inventory capacity, while an operation waits for approval or external review.
Sagas decompose long-running transactions into steps with compensating actions~\cite{garciamolina1987sagas}; \sys{} is complementary, treating escalation as a split-phase governed operation whose terminal commit must either consume a reservation or revalidate policy state.
Stateful serverless systems such as Beldi~\cite{beldi2020} similarly show that stateless function execution needs transactional and logging mechanisms; \sys{} applies that lesson to governance decisions over mutable policy state.

\paragraph{Audit, agent coordination, and evaluation.}
Database provenance explains why query results arise from underlying data~\cite{buneman2001why,cheney2009provenance}.
\sys{} has a narrower audit goal: for each terminal governed operation, it records the policy version, policy-state reads, decision, and effect outcome for the same logical operation.
Recent agent systems study concurrency over shared state; S-Bus is close because it reconstructs agent read sets and applies optimistic concurrency control~\cite{sbus2026}, while \sys{} focuses on policy-authored invariants and decision-effect atomicity.
Benchmarks such as \(\tau\)-bench, AgentDojo, and ToolEmu evaluate tool-using agents and risky tool behavior~\cite{taubench2024,agentdojo2024,toolemu2023}; our evaluation remains controlled so concurrency, conflict rate, approval delay, and \pss{} violations can be isolated directly.

\section{Conclusion}
\label{sec:conclusion}

Stateful governance turns authorization into a consistency problem.
When policies depend on mutable shared policy state, an allow decision is not a timeless capability.
It is valid only for the policy state on which it was evaluated and for the effect it authorizes.

The \sys{} runtime addresses this problem by keeping policies first-class while coordinating the policy-state facts that make their decisions sound.
Policy authors write bounded policies over certified policy-state views.
Policy-state providers expose the corresponding trusted effects and enforcement mechanisms.
The \sys{} coordinator maps the resulting dependencies to a suitable enforcement mechanism and records an audit trail for the same logical operation.

The broader point is that concurrent agentic systems need governance abstractions that are both programmable and consistency-aware.
Hand-written transactions can enforce fixed policies, and global serialization can provide correctness, but neither is a satisfactory abstraction for evolving, concurrent governance.
Together, \pss{} and the \sys{} runtime architecture are a step toward treating stateful governance as a systems problem with an explicit correctness condition and implementation boundary.

\begin{acks}
We thank Chunwei Liu, Hanshen Xiao, and Jianguo Wang for helpful discussions.
\end{acks}

\bibliographystyle{ACM-Reference-Format}
\bibliography{refs}

\appendix
\section{Correctness Proof}
\label{app:correctness-proof}

\paragraph{Proof of Theorem~\ref{thm:pss-scoped-enforcement}.}
Fix a terminal history $\History$ produced by a runtime satisfying the theorem assumptions.
For each terminal operation $i$, choose a serialization point inside its execution interval: for an allowed operation, the point at which the selected enforcement mechanism commits the governed effect; for a denied operation, the point at which the runtime records the denial.
Let $\Serial(\History)$ be the total order induced by these serialization points, with any exact ties broken by the runtime's fixed tie breaker.
Because each serialization point lies between the operation's begin and terminal events, this order respects the real-time order of $\History$.

We next show that each terminal decision is legal at its position in this serial order.
Consider a terminal operation $i$ and the policy state immediately before $i$ in $\Serial(\History)$.
By contract soundness, every policy-state fact read by $i$'s policy evaluation is covered by $i$'s logical scopes, and every policy-state fact read or written by any effect is covered by that effect's scopes.
Thus, any concurrent effect that could change a fact read by $i$'s policy has an overlapping logical scope with $i$.
The enforcement assumption says that overlapping scopes are serialized, reserved, or validated before a terminal decision is recorded and before any allowed effect commits.
Validation checks that the relevant scoped state is unchanged, reserved for the operation, or continuously protected by scoped enforcement.
Therefore, at $i$'s serialization point, no unvalidated concurrent effect has changed a policy-state fact on which $i$'s decision depends.
The certified policy-state view used by $i$ is consequently the same view that would be obtained from the policy state immediately before $i$ in $\Serial(\History)$.
Since policies are well typed, bounded, and pure over the request and certified view values, the recorded allow or deny decision for $i$ is exactly the decision produced by evaluating the applicable policies at that serial position.
If the decision is allow, the governed effect commits at that position; if the decision is deny, the operation produces no governed effect.

Pending operations do not enter $\History$ while pending.
When such an operation later becomes terminal, it does so only after the runtime either revalidates the relevant policy state or consumes a reservation whose protected scope set was established by the same sound-contract discipline.
For reservation mode, the proof relies on the escrow invariant stated by the provider contract: outstanding reservations are excluded from capacity available to other operations, and consuming a reservation retires that reserved capacity exactly once.
Thus two pending operations cannot both consume the same budget or inventory capacity.
It can therefore be treated as an allowed or denied terminal operation at its resolution point.

Finally, $\Serial(\History)$ contains exactly the same terminal operations as $\History$.
Allowed operations apply the same governed effects at their serialization points, and denied operations apply none.
Because the serial order follows the points at which these effects commit or denials are recorded, applying the effects in $\Serial(\History)$ yields the same visible governed effects and the same final policy state as the runtime history.
All five clauses of Definition~\ref{def:pss} hold, so $\History$ satisfies \pss{}.

\section{Policy Language Details}
\label{app:policy-language-details}

The main text relies only on two properties of the policy interface: policy programs are bounded and every certified policy-state view call can be mapped to logical scopes.
The prototype realizes this interface with a small DSL.

\paragraph{Policy form.}
A policy names the governed action, contains ordered deny or escalate rules, and ends with a default allow rule:

\begin{lstlisting}[style=masugatepolicy]
policy team_transfer_guard on transfer {
  deny insufficient_funds when
    accounts.balance(principal.id) < args.amount;

  deny daily_team_budget when
    budget.spent(principal.team, 24h)
      + args.amount > budget.limit(principal.team);

  allow otherwise;
}
\end{lstlisting}

The expression language is first-order and side-effect free.
It supports typed request and principal paths, simple arithmetic and Boolean expressions, and calls to registered certified policy-state views.

\paragraph{Static checks.}
Action arguments and principal attributes are typed by the effect contract, and policy-state views have registered function types.
The compiler rejects unregistered views, nested view calls, missing default rules, ill-typed expressions, mutation, loops, and arbitrary callbacks.
The implementation also caps the number of view calls per policy.

\begin{table}[h]
  \caption{Policy DSL static checks and dependency contributions.}
  \label{tbl:dsl-static-semantics}
  \begin{tabular}{p{0.30\columnwidth}p{0.30\columnwidth}p{0.30\columnwidth}}
    \toprule
    Form & Static rule & Dependency contribution \\
    \midrule
    \texttt{args.x} & \texttt{x} in effect arguments & none \\
    \texttt{principal.x} & \texttt{x} in principal schema & none \\
    \texttt{e1 + e2} & both operands \texttt{Int} & union of operands \\
    \texttt{e1 and e2} & both operands \texttt{Bool} & union of operands \\
    \texttt{Q(e1,...,en)} & registered bounded view & one or more logical scopes from \texttt{Q}'s resolver \\
    rule condition & expression has type \texttt{Bool} & scopes of the condition \\
    \bottomrule
  \end{tabular}
\end{table}

\paragraph{Operational semantics.}
Rules are evaluated in order.
The first rule whose condition evaluates to true determines the decision.
If no deny or escalate rule fires, the trailing \texttt{allow otherwise} rule produces an allow decision.
Each view call is evaluated through the registered resolver in a provider-owned session and records the returned value plus audit metadata.
Policy evaluation therefore produces both a decision and the dependency trace used by the runtime.

\section{Policy-Evolution Fixtures}
\label{app:policy-evolution-fixtures}

RQ4 measures source changes over a small family of policy-evolution fixtures.
The \sys{} side changes policy programs and, when a new certified policy-state view is needed, a provider contract.
The expert baseline embeds the same policy logic inside trusted provider code.
Listings~\ref{lst:rq4-masugate-policies} and~\ref{lst:rq4-expert-programs} show the policy and expert programs used for the RQ4 source-delta counts; filename comments are added only to identify the fixture variants.

\begin{lstlisting}[style=masugatepolicy,basicstyle=\ttfamily\scriptsize,caption={MasuGate policy programs used in RQ4.},label={lst:rq4-masugate-policies}]
# base/masugate_policy.pvl
policy team_transfer_guard on transfer {
  deny insufficient_funds when
    accounts.balance(principal.id) < args.amount_cents;
  deny daily_team_budget when
    ledger.sum_sent_by_team(principal.team, 24h) + args.amount_cents > 100000;
  allow otherwise;
}

# per_agent/masugate_policy.pvl
policy team_transfer_guard on transfer {
  deny insufficient_funds when
    accounts.balance(principal.id) < args.amount_cents;
  deny daily_team_budget when
    ledger.sum_sent_by_team(principal.team, 24h) + args.amount_cents > 100000;
  deny per_agent_limit when
    ledger.sum_sent_by_agent(principal.id, 24h) + args.amount_cents > 25000;
  allow otherwise;
}

# approval/masugate_policy.pvl
policy team_transfer_guard on transfer {
  deny insufficient_funds when
    accounts.balance(principal.id) < args.amount_cents;
  deny daily_team_budget when
    ledger.sum_sent_by_team(principal.team, 24h) + args.amount_cents > 100000;
  deny missing_approval when
    approvals.count(args.request_id) < 2;
  allow otherwise;
}

# risk/masugate_policy.pvl
policy team_transfer_guard on transfer {
  deny insufficient_funds when
    accounts.balance(principal.id) < args.amount_cents;
  deny daily_team_budget when
    ledger.sum_sent_by_team(principal.team, 24h) + args.amount_cents > 100000;
  deny risky_principal when
    risk.score(principal.id) > 80;
  allow otherwise;
}

# window_7d/masugate_policy.pvl
policy team_transfer_guard on transfer {
  deny insufficient_funds when
    accounts.balance(principal.id) < args.amount_cents;
  deny weekly_team_budget when
    ledger.sum_sent_by_team(principal.team, 7d) + args.amount_cents > 100000;
  allow otherwise;
}

# second_action/masugate_policy.pvl
policy team_transfer_guard on transfer {
  deny insufficient_funds when
    accounts.balance(principal.id) < args.amount_cents;
  deny daily_team_budget when
    ledger.sum_sent_by_team(principal.team, 24h) + args.amount_cents > 100000;
  allow otherwise;
}

policy team_withdraw_guard on withdraw {
  deny insufficient_funds when
    accounts.balance(principal.id) < args.amount_cents;
  deny daily_team_budget when
    ledger.sum_sent_by_team(principal.team, 24h) + args.amount_cents > 100000;
  allow otherwise;
}
\end{lstlisting}

\begin{lstlisting}[style=masugatepython,basicstyle=\ttfamily\scriptsize,caption={Expert baseline programs used in RQ4.},label={lst:rq4-expert-programs}]
# base/expert_resource.py
def transfer_with_policy(tx, request):
    sender = tx.account(request.principal.id)
    spent = tx.sum_sent_by_team(request.principal.team, hours=24)
    if sender.balance_cents < request.amount_cents:
        return deny("insufficient_funds")
    if spent + request.amount_cents > 100000:
        return deny("daily_team_budget")
    return tx.transfer(request)

# per_agent/expert_resource.py
def transfer_with_policy(tx, request):
    sender = tx.account(request.principal.id)
    spent = tx.sum_sent_by_team(request.principal.team, hours=24)
    agent_spent = tx.sum_sent_by_agent(request.principal.id, hours=24)
    if sender.balance_cents < request.amount_cents:
        return deny("insufficient_funds")
    if spent + request.amount_cents > 100000:
        return deny("daily_team_budget")
    if agent_spent + request.amount_cents > 25000:
        return deny("per_agent_limit")
    return tx.transfer(request)

# approval/expert_resource.py
def transfer_with_policy(tx, request):
    sender = tx.account(request.principal.id)
    spent = tx.sum_sent_by_team(request.principal.team, hours=24)
    approvals = tx.approval_count(request.request_id)
    if sender.balance_cents < request.amount_cents:
        return deny("insufficient_funds")
    if spent + request.amount_cents > 100000:
        return deny("daily_team_budget")
    if approvals < 2:
        return deny("missing_approval")
    return tx.transfer(request)

# risk/expert_resource.py
def transfer_with_policy(tx, request):
    sender = tx.account(request.principal.id)
    spent = tx.sum_sent_by_team(request.principal.team, hours=24)
    risk = tx.risk_score(request.principal.id)
    if sender.balance_cents < request.amount_cents:
        return deny("insufficient_funds")
    if spent + request.amount_cents > 100000:
        return deny("daily_team_budget")
    if risk > 80:
        return deny("risky_principal")
    return tx.transfer(request)

# window_7d/expert_resource.py
def transfer_with_policy(tx, request):
    sender = tx.account(request.principal.id)
    spent = tx.sum_sent_by_team(request.principal.team, days=7)
    if sender.balance_cents < request.amount_cents:
        return deny("insufficient_funds")
    if spent + request.amount_cents > 100000:
        return deny("weekly_team_budget")
    return tx.transfer(request)

# second_action/expert_resource.py
def transfer_with_policy(tx, request):
    sender = tx.account(request.principal.id)
    spent = tx.sum_sent_by_team(request.principal.team, hours=24)
    if sender.balance_cents < request.amount_cents:
        return deny("insufficient_funds")
    if spent + request.amount_cents > 100000:
        return deny("daily_team_budget")
    return tx.transfer(request)

def withdraw_with_policy(tx, request):
    sender = tx.account(request.principal.id)
    spent = tx.sum_sent_by_team(request.principal.team, hours=24)
    if sender.balance_cents < request.amount_cents:
        return deny("insufficient_funds")
    if spent + request.amount_cents > 100000:
        return deny("daily_team_budget")
    return tx.withdraw(request)
\end{lstlisting}

\end{document}